Title

# Electrically driven plasmon-polaritonic bistability in Dirac electron tunneling transistors

Authors

Shuai Zhang[1]†*, Yang Xu[2]†, Junhe Zhang[1], Dihao Sun[1], Yinan Dong[1], Matthew Fu[1], Takashi Taniguchi[3], Kenji Watanabe[4], Cory Dean[1], Monica Allen[5], Jeffery Allen[5], F. Javier García de Abajo[6,7], Antti J. Moilanen[2], Lukas Novotny[2]*, D.N. Basov[1]*

Affiliations

[1]Department of Physics, Columbia University, New York, NY, 10027, USA.

[2]Photonics Laboratory, ETH Zürich, Zürich, Switzerland.

[3]Research Center for Materials Nanoarchitectonics, National Institute for Materials Science, 1-1 Namiki, Tsukuba 305-0044, Japan.

[4]Research Center for Electronic and Optical Materials, National Institute for Materials Science, 1-1 Namiki, Tsukuba 305-0044, Japan.

[5]Air Force Research Laboratory, 101 W. Eglin Blvd, Bldg. 832, Eglin AFB, FL, 32542, USA.

[6]ICFO-Institut de Ciencies Fotoniques, The Barcelona Institute of Science and Technology, Castelldefels (Barcelona), 08860, Spain.

[7]ICREA-Institució Catalana de Recerca i Estudis Avançats, Passeig Lluís Companys 23, 08010, Barcelona, Spain.

*Corresponding author email: szhangphysics@gmail.com, lukas.novotny@ethz.ch, db3056@columbia.edu.
†These authors contributed equally: Shuai Zhang, Yang Xu

Abstract

Bistability—two distinct stable states under identical parameters—is not only a fundamental physical concept but also of importance in practical applications. While plasmon-polaritonic bistability representing history-dependent stable states within plasmonic systems has been theoretically predicted, it has yet to be demonstrated experimentally due to challenges in realizing suitable nonlinearity at feasible electric-field strengths. Here, we report the experimental observation of electrically driven plasmon-polaritonic bistability in graphene/hexagonal-boron-nitride/graphene tunneling transistors, achieved through momentum-conserving resonant tunneling of Dirac electrons. Using a small twist angle between graphene layers, we engineered devices exhibiting both electronic and plasmon-polaritonic bistability. This bistable plasmonic behavior can be precisely tuned through load resistance and electrostatic gating. Our findings open new pathways for exploring nonlinear optical and electronic phenomena in van der Waals heterostructures and mark a significant advance in nanoplasmonics, with potential applications in optical memory, sensing, and optoelectronic switching.

**MAIN TEXT**

Bistability—a key feature of nonlinear systems—is central to functional devices, such as those involving logic operations and information storage [1-4]. The search for intrinsic and robust bistable states has prompted extensive studies of van der Waals (vdW) heterostructures formed by stacking various constituents with atomic-layer precision [5,6]. Heterostructures based on graphene and other two-dimensional crystals (e.g., hexagonal boron nitride (hBN)) have demonstrated fascinating electrical and optoelectronic properties, including the presence of tunneling transistors [7], negative differential conductance [8-10], and tunneling electron-induced light emission [11]. Notably, bistable electronic states have been reported in heterostructures with graphene, including bistable superconducting and correlated insulating states in graphene/hBN heterostructures [12] and Josephson tunneling with hysteretic features in twisted graphene [12,13]. In addition to these rich and intriguing electrical properties [14], graphene possesses exciting optical properties that include a high optical nonlinearity and the presence of tunable, highly confined, long-lived plasmon polaritons [15,16]. However, bistable optical or plasmonic states in graphene or its heterostructures remain elusive.

Here, we use graphene/hBN/graphene (Gr/hBN/Gr) tunneling transistors, with a small twist angle between graphene layers, to realize bistability in electronic properties that drives bistability in plasmon-polaritonic properties. We demonstrate that negative differential conductance (NDC), arising from resonant tunneling of Dirac electrons, enables electrical transport bistability. Furthermore, we show that resonant tunneling can be exploited to manipulate plasmon polaritons in graphene tunneling transistors. Notably, we observe electrically driven plasmon-polaritonic bistability for the first time and demonstrate that this effect can be effectively controlled through load resistance and gate voltage.

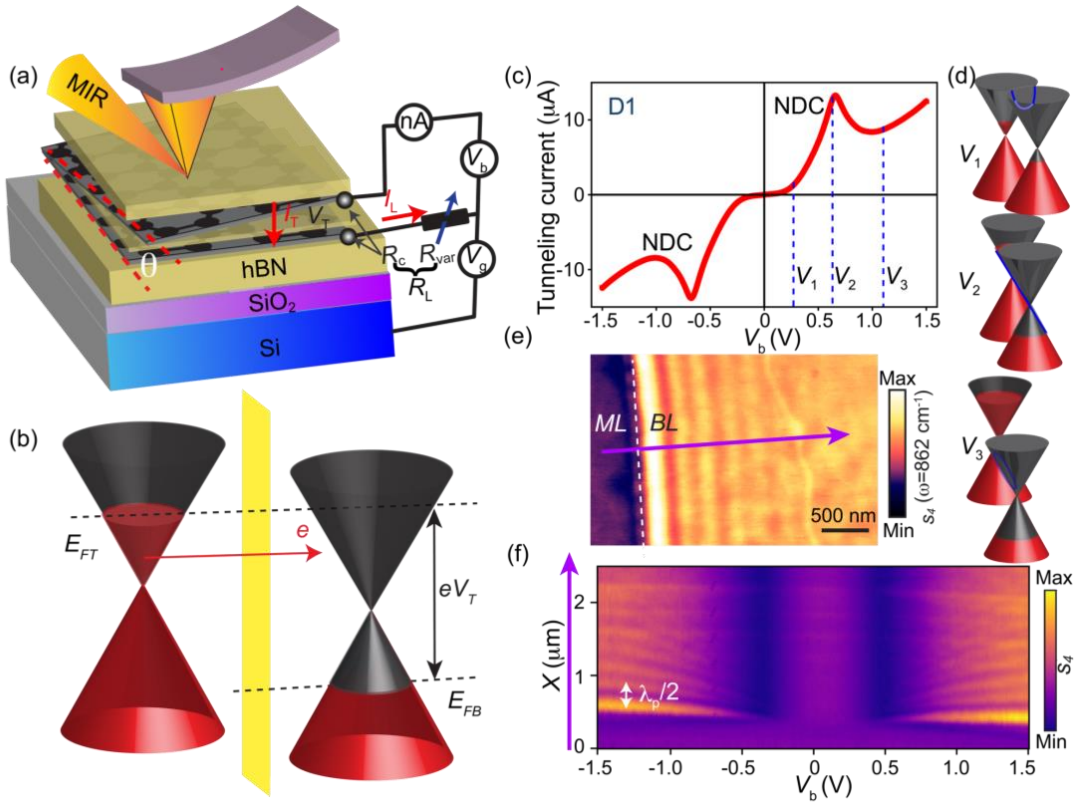

FIG. 1. **Current-voltage characteristics and plasmonic response in graphene/hBN/graphene (Gr/hBN/Gr) resonant tunnel junctions.** (a) Schematic of a twist-controlled Gr/hBN/Gr tunnel junction with a bottom gate and the experimental setup. There is a small twist angle $\theta \sim 1°$ between the two graphene electrodes. Under mid-infrared (MIR) illumination, graphene plasmons are visualized with scattering-type scanning near-field optical microscopy. The load resistance $R_L$ is the sum of the variable resistance $R_{var}$ and the contact resistance $R_c$. The gate voltage $V_g$ is applied to the silicon backgate. (b) Electronic band diagram of the Gr/hBN/Gr tunnel junction when a positive bias voltage $V_T$ is applied to the bottom graphene electrode. Electrons can tunnel through a thin barrier of hBN both elastically and inelastically. The electrostatic doping of graphene layers is controlled by the bias voltage and the gate voltage $V_g$. (c) Experimental current-voltage (I-V) characteristics obtained for a representative device. Symmetric negative differential conductance (NDC) arises from the resonant tunneling process. (d) Fermi energies and Dirac cone alignments of two graphene electrodes at three characteristic bias regimes. (e) Nano-IR scattering $s_4$ image of device 1 (D1) under a bias voltage of $V_b = +1.5$ V. The boundary between monolayer graphene (ML) and the tunnel junction with bilayer graphene (BL) is marked by a white dashed line. (f) Plasmonic dispersion of the resonant tunneling transistor as a function of bias voltage, constructed from line scans along $X$ acquired at a fixed position marked by a purple line in (e). The photon energy in (e) and (f) is $\omega = 862 \text{ cm}^{-1}$.

*Device architecture*—A schematic of the resonant tunneling transistor and experimental setup is shown in Fig. 1(a). We used the dry transfer method to assemble the heterostructures from individual flakes prepared by mechanical exfoliation. The tunnel junction is created with an atomically thin hBN barrier separating two graphene layers. The two graphene layers are twisted by ~1° relative to each other. This angle is critical for the present study, as detailed below. We investigated Gr/hBN/Gr structures in which the thickness of the hBN tunneling barrier ranges from 3 to 5 monolayers, ensuring a considerable tunneling current density. The tunnel junctions are encapsulated by two additional thicker hBN flakes on both the top and bottom sides to preserve the high electrical quality of the devices. The entire stack is assembled on a doped silicon substrate with a 285-nm-thick oxidized layer. The silicon serves as a backgate electrode with an applied gate voltage of $V_g$. We added a variable resistor in series with the tunnel junction. The total resistance of the circuit is produced by the tunnel junction resistance $R_T$, the variable resistance $R_{var}$, and the contact resistance $R_c$ of the graphene sheets. The load resistance $R_L$ is given by $R_L = R_{var} + R_c$; the contact resistance $R_c$ is ignored for simplicity when specifying the load resistance, unless otherwise stated. Therefore, the bias voltage applied to the devices, $V_b$, is divided as $V_b = V_T + V_L$, where $V_T$ is the tunnel junction voltage and $V_L$ is the voltage across the load resistor. The voltage $V_T$ results in a chemical potential difference between the top and bottom graphene electrodes, generating a tunneling current, as shown in Fig. 1(b).

The working principle of the tunneling device is illustrated in Fig. 1(b). The backgate controls the doping level of both the top and bottom graphene layers, due to the weak screening of graphene [7]. For simplicity, we first present data obtained without a backgate. By applying a voltage $V_T$ across the tunnel junction, one creates the chemical potential difference associated with $V_T$ between the two graphene layers. In addition, the voltage $V_T$ tunes the Fermi energies of the top (T) and bottom (B) graphene, $E_{FT}$ and $E_{FB}$, with respect to the charge neutrality point, due to the quantum capacitance of graphene [17]. We note that the use of a thin hBN spacer between the graphene layers results in a significant quantum capacitance [17]. Consequently, the energy difference between the two charge neutrality points, $\phi$, is generally substantially smaller than $V_T$ and is given by $\phi = eV_T - 2E_F$. The resulting energy band alignment governs the electron tunneling between the two graphene layers.

*Negative differential conductance*—We first investigate twist-controlled resonant tunneling in our devices. Figure 1(c) shows the tunneling current measured in a representative Gr/hBN/Gr device as a function of the applied voltage $V_b$. The data feature two current peaks, corresponding to two current-voltage (I-V) regimes that exhibit negative differential conductance (NDC). The locations of these current peaks depend on the twist angle between the graphene layers and are rooted in momentum-conserving resonant tunneling [9,18-20]. For a small twist angle $\theta$ between the two graphene lattices, the Brillouin zones of graphene accordingly rotate by $\theta$. This rotation results in an in-plane Dirac cone displacement, $\Delta \boldsymbol{K} \approx \hat{\boldsymbol{z}} \times \overrightarrow{\boldsymbol{\Gamma K}} \theta$, where $\hat{\boldsymbol{z}}$ is the unit vector perpendicular to the graphene plane, and $\Gamma$ and $K$ are the center and corner of the graphene Brillouin zone, respectively. With an applied voltage of $V_b$, the Dirac cones shift vertically, as illustrated in Fig. 1(d). At the specific bias voltage where $\phi = \hbar v_F \Delta K$, the Dirac cones of the two individual graphene layers intersect along a line (middle panel of Fig. 1(d)). Under this latter condition, conservation of momentum and energy is satisfied for a large fraction of states, resulting in a surge in the tunneling rate. Consequently, the tunneling current peaks at this specific bias voltage. Tunneling simulations with the assumption of momentum conservation based on Bardeen's tunneling theory reproduce the experimental data well, as shown in Supplementary Notes 1-2 and Figs. S1-S4.

*Plasmon excitation*—Now we turn to the plasmonic response of doped tunneling devices under a bias voltage. The voltage $V_T$ between the two graphene electrodes gives rise to two concurrent effects [17]: 1) $V_T$ offsets the Fermi energy difference between the two graphene layers; and 2) $V_T$ produces doping of the graphene layers because a displacement field, $\varepsilon \phi / d$, is formed in the tunnel junction, where $d$ and $\varepsilon$ are the thickness and permittivity of the hBN tunneling barrier, respectively. This carrier doping results in tunable plasmon-polariton excitations of the double graphene layers, including acoustic and optical modes [21-23].

The plasmon polaritons with high momentum [24] are investigated by focusing mid-infrared light onto the sharp tip of a scattering-type scanning near-field optical microscope (s-SNOM). The tip-launched plasmon polaritons are reflected by the device edge and, upon completing a round trip, produce interference fringes in the measured scattering signal as a function of tip-edge distance [25-27], as shown in Fig. 1(e). The excitation frequency is selected based on the simulated plasmonic dispersion, which is shown in Supplementary

Fig. S6. We note that our infrared scattering data are dominated by the optical plasmon mode, as the acoustic mode possesses a higher momentum and cannot be efficiently coupled to the tip (Supplementary Note 4 and Fig. S6). The plasmon-polariton dispersion is evaluated along the line indicated by a purple arrow in Fig. 1(e) for different bias voltages, as shown by a two-dimensional false-color map in Fig. 1(f). In the course of this data acquisition, we repeatedly scanned the tip of the s-SNOM along the same line while gradually sweeping the bias voltage and recording the scattering signal. As shown in Fig. 1(f), plasmon-polariton fringes emerge at higher bias voltages. Furthermore, the fringe spacing increases with increasing bias voltage, indicating that the dispersion is bias voltage-dependent. (See additional data and dispersion analysis in Supplementary Note 4 and Fig. S7).

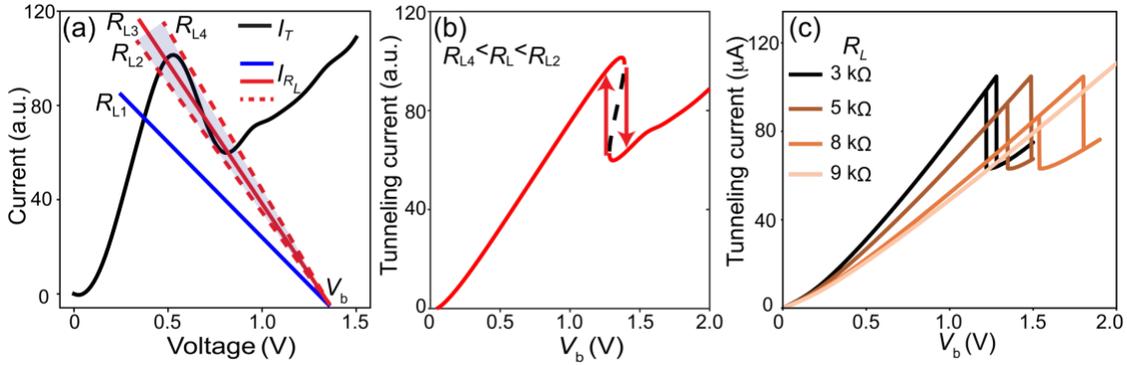

**FIG. 2. Electrical transport hysteresis arising from tunneling bistability**. (a) Schematic illustrating the realization of bistable states in a resonance tunneling device. The black curve represents a simulated resonant tunneling I-V curve exhibiting negative differential conductance. The blue and red straight lines (referred to as load lines) denote the I-V characteristics for various resistances $R_L = R_{L1}, R_{L2}, R_{L3}$, and $R_{L4}$. The slopes of these load lines are given by $-1/R_L$. Current in the device is governed by the nonlinear equation $I_T(V_T) = I_L(V_b - V_T)$, where $I_T$, $V_T$, and $I_L$ are the tunneling current, the voltage between the graphene layers, and the current across the load resistor, respectively. When there are three points of intersection between the tunneling line and the load line ($R_{L4} < R_L < R_{L2}$), a bistable I-V behavior is displayed. (b) Simulated bistable I-V characteristics with load resistance within the bistable range. (c) Measured I-V curves with different load resistances, demonstrating the bistable behavior.

*Bistable I-V states*—A negative differential conductance can be exploited to generate multivalued states that produce bistability. The bistable I-V characteristics of the device are governed by the nonlinear equation $I_T(V_T) = I_L(V_b - V_T)$, where $I_T$, $I_L$, $V_T$, and $V_b$ are the tunneling current, the current through the load resistor, the voltage between the two graphene tunneling electrodes, and the voltage applied to the device, respectively. The bistable regimes implied by this equation are illustrated in Fig. 2(a). In this schematic, the load lines, marked by straight lines, represent the response of varying load resistances in the circuit. The load lines are denoted by $I_L = -\frac{V_T - V_b}{R_L}$. For specific loads within the gray shaded regime, as exemplified by the red solid line, there are three points of intersection between the tunneling curve (black curve) and the load line, indicating three solutions to the transport equation. That is, with such critical values of the load resistances, there are three I-V states at a given bias voltage, as shown by the simulated I-V curve in Fig. 2(b). However, one of these three states (marked by a dashed line) is not stable: a slight perturbation will cause it to converge to one of the other two stable equilibrium states, and thus, bistable I-V characteristics can be formed. A bifurcation—a sudden change in the qualitative behavior of a system caused by a small change in a control variable—occurs at the thresholds for three states, marked by red dashed lines in Fig. 2(a). We note that similar intrinsic bistable tunneling was first predicted by J. F. Rodriguez-Nieva *et al.* in a trilayer graphene architecture [28]. This seminal theoretical work provided a foundational framework for understanding transport bistability driven by twist-angle-controlled resonant tunneling. By revealing the critical role of momentum-conserving tunneling in enabling intrinsic multistable states, it laid essential groundwork for subsequent experimental efforts. Motivated by these insights, we adapted and optimized our device architecture to realize and extend these predictions in a practical platform, as detailed above.

As illustrated in Fig. 2(a), the bistable I-V behavior depends on the slope of the load line, which is determined by the load resistance. Figure 2(c) shows experimental I-V traces for our device, measured with different load resistances. Hysteresis—a signature of bistable states—can be observed over a range of load resistances. The position and width of the hysteresis loop depend on the load resistance. In addition, as bias voltages are tuned, the states switch abruptly, as shown schematically by red arrows in Fig. 2(b). The simulated I-V bistability is shown in Supplementary Note 3 and Fig. S5.

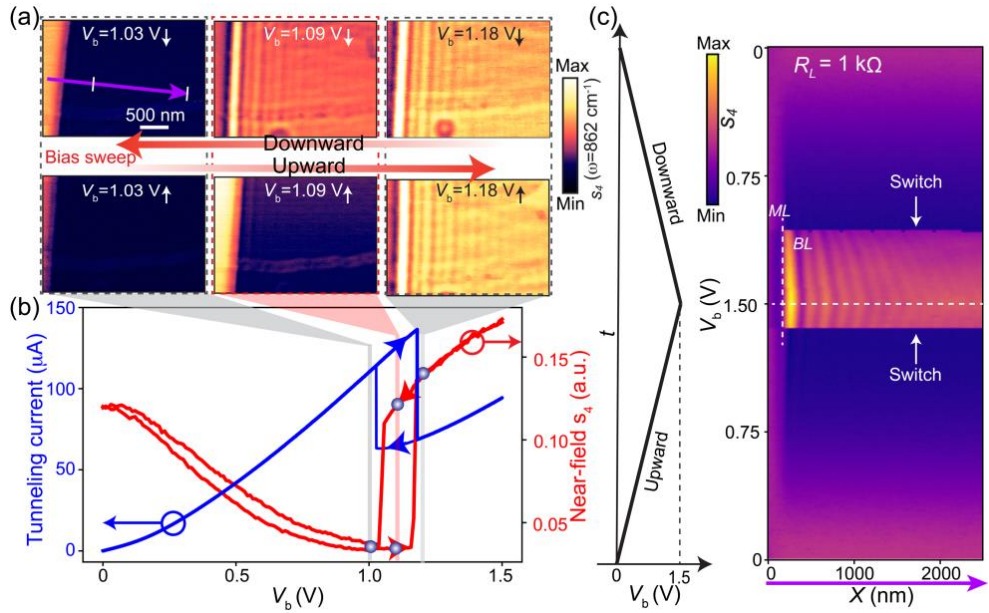

**FIG. 3. Plasmon-polaritonic bistability.** (a) Nano-IR images acquired when the bias voltage $V_b$ is swept up (bottom panels) and down (top panels). (b) Correlation between bistable tunneling behavior (blue) and bistable plasmon-polaritonic response (red), both of which exhibit hysteretic behaviors. The near-field $s_4$ is extracted by averaging along a line segment with boundaries marked by two white lines in (a). (c) Bistable plasmonic response observed in the 2D nano-IR scattering data, $s_4$. An abrupt switch between plasmonic states (indicated by white arrows) is observed at different bias voltages in upward and downward sweeping along a common scanning line $X$ (purple line in (a)) with a 0 to 1.5 V voltage excursion (see left part of the panel). All the data were acquired on device 2 (D2).

*Bistable plasmonic states*—The bistable I-V characteristics described above give rise to discontinuous tunnel voltages and, hence, to discontinuous doping of the graphene layers. Thus, at a specific voltage, the carrier density abruptly switches from one value to another, potentially leading to two distinct plasmon-polaritonic responses. To experimentally confirm the bistable plasmon-polaritonic states, we acquired nano-IR scattering images in the course of both upward and downward voltage sweeps. Representative nano-IR images in the bottom panels of Fig. 3(a) were acquired by increasing the voltage, while images in the top panels were collected when the applied voltage was swept down. Additional images at varying bias voltages are shown in Supplementary Fig. S9. Notably, the two nano-IR images collected at the same voltage (middle column) display distinct features. That is, the plasmon-polaritonic response is determined by the voltage sweep history and exhibits hysteretic behavior. The observed concomitant hysteretic loops in the tunneling current and near-field light scattering are presented in Fig. 3(b). The hysteresis arising from intrinsic bistability is pronounced and consistently appears alongside the sharp state switch. In contrast, extrinsic hysteresis is minor and, in some cases, barely discernible (Supplementary Note 5 and Fig. S10). In the hysteresis regime, the near-field intensity exhibits a pronounced contrast between the two bistable states; this arises primarily from the different coupling efficiencies between plasmon-polaritons and the near field of the atomic force microscope tip [29]. The polaritonic quality factor remains nearly constant in the hysteresis regime, as evidenced in Supplementary Fig. S8, and varies only marginally during the bistable state transition.

To more systematically illustrate the bistable plasmon-polaritonic response, we investigated the plasmonic response as a function of bias voltage. We recorded nano-IR scattering line scans for varying bias voltages, similar to those in Fig. 1(f). As shown in Fig. 3(c), a bistable plasmon-polaritonic behavior analogous to the electronic one is clearly observed in the optical signal: the plasmonic response abruptly switches at critical voltages, indicated by white arrows. In addition, the plasmonic responses at the upward and downward voltages are not the same, exhibiting asymmetric (or hysteretic) features. To the best of our knowledge, the data in Fig. 3 represent the first demonstration of plasmon-polaritonic bistability in a vdW heterostructure. In addition, we investigated the tunability of plasmon-polaritonic bistability, as detailed in Appendix A.

Although transport bistability has been observed in conventional semiconductors such as double-barrier quantum wells [29], plasmonic bistability in these structures has not been reported. In contrast, the low doping density required to substantially change the plasmonic behavior of graphene allows us to simultaneously observe both electrical and plasmonic bistability in twisted graphene-graphene tunneling devices. In addition, the bistable state switching time is controlled by the tunable RC time constant and can be optimized by choosing barrier materials with an appropriate thickness to tune the resistance and capacitance of the graphene tunneling transistors. The roadmap toward fast bistable switching and the present obstacles are detailed in Supplementary Note 5. For example, using transition metal dichalcogenides as tunneling barrier materials will dramatically increase the tunneling conductance [30,31], thereby resulting in a higher operation frequency, which can potentially reach the THz regime. It is noteworthy that THz emission has been observed in similar devices [32], suggesting that the underlying mechanism could be correlated with the resonant tunneling and bistable states reported here. Plasmonic bistable states can also control the optoelectrical response; this holds potential for the implementation of thermal/electric sensors. We demonstrate these latter capabilities by experimentally controlling the photothermoelectric current though the device bistability (Supplementary Note 5 and Fig. S12).

*Discussion*—Optical and plasmonic bistabilities reported in the literature were typically rooted in nonlinear optical conductivity as a function of the applied electric field [33,34]. The concept of optical bistability has expanded significantly. It now broadly encompasses phenomena driven by various other nonlinear mechanisms, such as nonlinear mechanical responses [35,36], heat- or light illumination-induced phase transitions [37,38], and nonlinear exciton responses driven by carrier injection or electrical fields in quantum wells [3,39]. To avoid any ambiguity in terminology, we emphasize that our work demonstrates electrically driven plasmon-polariton bistability (for more details, see Supplementary Note 8). The plasmonic bistability of graphene, based on the nonlinear Kerr effect, has been postulated theoretically [40,41]. In practice, however, realization of the proposed nonlinear plasmonic bistability has been hindered by the high electric field that is required to enhance the nonlinear response. In contrast, the approach demonstrated in our work leverages resonant tunneling to tune the graphene optical conductivity, thereby creating plasmonic bistability. Specifically, the observed plasmons do not depend on the intensity of the electric field and are emerging in the linear optical regime.

Compared to purely optical bistability, plasmon polariton bistability—with nanoscale spatial confinement and strongly enhanced field strength [42,43]—offers advantages such as flexible integration [44], sensitive state readout [45], and improved energy efficiency [46,47] (more details on the role of plasmon polaritons can be found in Supplementary Note 9). The remarkably low levels of doping needed to substantially modify the plasmonic response of graphene could be leveraged to electrically store information with a small number of electrons (and also, therefore, a low level of energy consumption) per bit, compared with the $10^5$ electrons needed in current field-effect transistor (FET) technology [1,48]. Reading could be performed either electrically or optically/plasmonically. In this respect, future studies need to address the issue of the minimum device area required to sustain a robust plasmonic bistability behavior. The fundamental limit for the footprint is imposed by the plasmon wavelength of ~10 nm. This length scale implies that a single electron per bit should be sufficient for switching under typical operating voltages [46]. This five-orders-of-magnitude reduction in the number of electrons per bit, together with the low bias voltages used compared to those in microchips, foretell a substantial saving in power, which could benefit from optimization of tunnel junction parameters. Our demonstrated electrically controlled plasmonic bistable states and their electrical readout hold great promise for applications such as plasmonic switches, modulators, memory devices, and advanced optical interconnect technologies [49] (more details on applications can be found in Supplementary Note 10 and Fig. S13). Recently, bistable tunneling and its optical switching have been leveraged to build sensitive photodetectors, demonstrating the potential for infrared vision neuron applications [50].

*Summary and outlook*—we have demonstrated concomitant bistability in both the current-voltage characteristics and the plasmon-polaritonic response of Gr/hBN/Gr tunnel junctions. Depending on the input history, these devices settle into distinct electrical and plasmon-polaronic states under identical inputs. The long-sought-after plasmon-polaritonic bistability has been eventually realized thanks to the advent of van der Waals materials, their heterostructures, and advancements in nanoplasmonics, thus complementing earlier breakthroughs in magnon- and exciton-polaritonic bistability [2,3]. This intrinsic plasmon-polaritonic bistability arises from a combination of the momentum-governed nonlinear resonance tunneling with highly carrier-density-dependent two-dimensional plasmons. The established bistability mechanism is likely to be operational in a broad range of material systems, which exhibit resonant tunneling and electrostatic tunable plasmons [51,52]. This new mechanism contrasts with the previously predicted counterpart originating from Kerr nonlinearity [40]; the electrically driven mechanism interfaces the electrons and photons, providing a platform readily integrable with on-chip electronics and photonics [49]. Furthermore, our findings confirm that the bistability can be flexibly tuned through parameters such as twist angle, tunneling barrier, and load resistance. We envision that the multivalued stable states, manifested in both current-voltage characteristics and nanoplasmonic response, hold promise for compact logic and storage circuits [53], neuromorphic sensing [50], and optical computation [54]. Future work needs to explore plasmon-assisted electron tunneling [55,56] and the excitation of plasmon polaritons by tunneling electrons [57]. Electrically excited and driven plasmonic bistability eliminates the need for the bulky light sources and complex momentum-compensation structures required for plasmon excitation. Furthermore, the electrically driven approach potentially enables plasmon amplification [58,59].

# APPENDIX

## A. Tunable plasmon-polaritonic bistability

Bistability is governed by the load line and tunneling properties. Therefore, the bistable states can be controlled by the load resistance. I-V characteristics with various load resistances $R_L$ are shown in Fig. 4(a). The corresponding nano-IR data are illustrated in Figs. 4(b-d). In particular, Fig. 4(b) shows that the plasmonic response continuously evolves with the bias voltage. In striking contrast, with a larger load resistance, the plasmonic response exhibits an abrupt switch when the voltage is swept (Figs. 4(c,d), marked by white arrows). Concomitantly, asymmetric features with respect to the maximum bias voltage are observed (Figs. 4(c,d), marked by horizontal white dashed lines). To more clearly show the abrupt switch and hysteresis of bistable plasmon-polaritonic states, line cuts are taken from Figs. 4(b-d) and plotted in Figs. 4(e-g). Similar results are also observed in other devices, as shown in Supplementary Fig. S11.

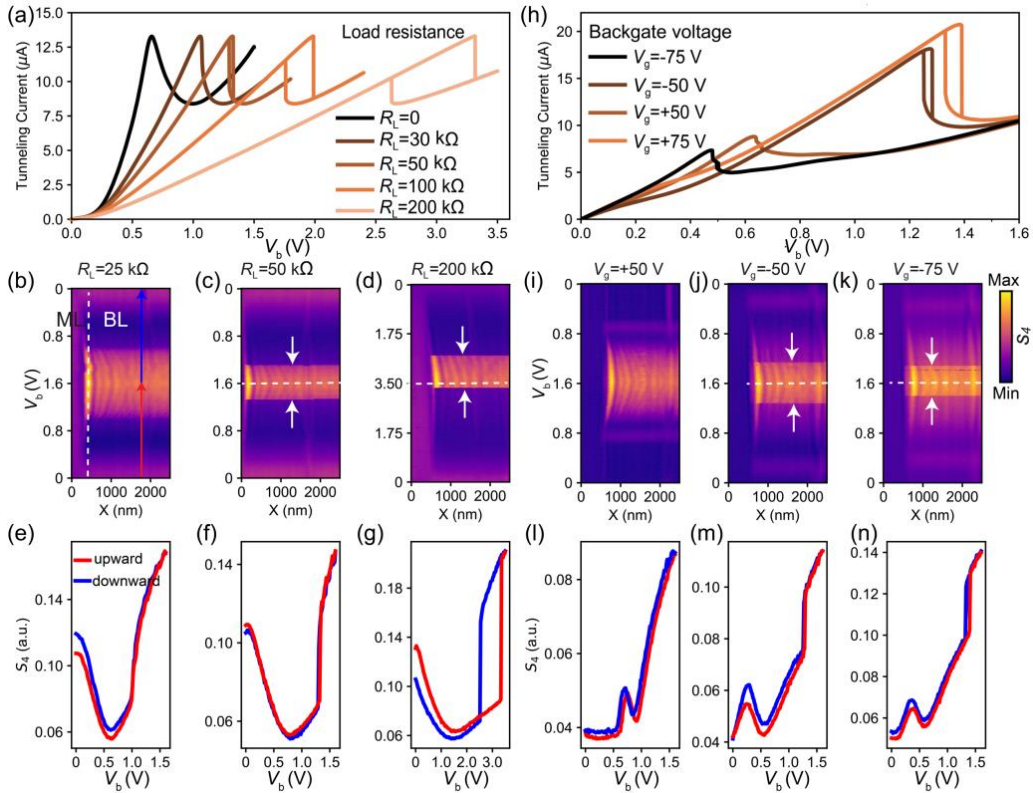

**FIG. 4. Plasmon-polaritonic bistability tuned via load resistance and backgate voltage.** (a) Measured I-V characteristics with various load resistances $R_L$ for a fixed backgate voltage $V_g = 0$. (b, c and d) Nano-IR scattering ($s_4$) line scans along a line perpendicular to the monolayer and double graphene boundary (see Fig. 1(e)), with different load resistances

(upper labels), following the same procedure and notation as in Fig. 3(c). (h) Measured I-V characteristics with various backgate voltages $V_g$ and a fixed load resistance $R_L = R_C$. (i, j, and k) Nano-IR scattering ($s_4$) line scans with different backgate voltages (upper labels). The measurement procedure is similar to that in (b-d). (e-g, l-n) Line profiles extracted from the corresponding nano-IR scattering line scans along the blue (downward $V_b$) and red (upward $V_b$) arrows indicated in panel (b). The emergence of plasmon-polaritonic hysteresis and the width of the hysteresis loop can be tuned by either the load resistance or the backgate voltage. All the data were acquired on D1.

The backgate doping can change the Fermi energy of both graphene electrodes [7], thereby modifying electron tunneling between the graphene monolayers. Since the plasmonic bistability is governed by resonant tunneling, the backgate is well suited to control the I-V characteristics and plasmon-polaritonic bistability. The I-V characteristics obtained by varying the backgate voltage $V_g$ are shown in Fig. 4(h). Polaritonic data collected with three representative backgate voltages are illustrated in Figs. 4(i-k). As expected from Fig. 4(h), Fig. 4(i) shows that the plasmonic response continuously evolves with the bias voltage. In contrast, in Figs. 4(j,k), the plasmon-polaritonic response exhibits an abrupt switching, marked again by white arrows. Linecuts taken from Figs. 4(i-k) and plotted in Figs. 4(l-n) clearly show bistable plasmon-polaritonic switching and hysteresis. We can also see that the width of the hysteresis loop can be efficiently tuned by the backgate voltage. Note that bistability can be optimized on demand not only by means of the load resistance and the backgate voltage but also via the twist angle between the two graphene layers, the tunneling barrier material, and the barrier thickness. Overall, bistability arises from the linear dispersion of Dirac electrons, the low electronic density of the states, and the unique plasmonic response. Therefore, the bistability reported in the present work is intrinsic and robust.

**Acknowledgments**

**Funding:** Bistability experiments at Columbia were supported by the University of Dayton Research Institute (UDRI) under contract SUB-25-000008. Research on graphene plasmons is supported by DOE-BESDE-SC0018426. The nano-IR measurement is supported as part of Programmable Quantum Materials, an Energy Frontier Research Center funded by the US Department of Energy (DOE), Office of Science, Basic Energy Sciences (BES), under award DE-SC0019443. D.N.B. is the Moore investigator in quantum materials, EPIQS GBMF9455. Research at ETH was supported by the Swiss National Science Foundation (grant 200020_192362/1) and the ETH Grant SYNEMA ETH-15 19-1. A.J.M. acknowledges financial support from the ETH Zürich Postdoctoral Fellowship programme, and Y.X. acknowledges the use of the cleanroom facilities at the FIRST Center for Micro- and Nanoscience at ETH Zürich. K.W. and T.T. acknowledge support from the JSPS KAKENHI (Grant Numbers 21H05233 and 23H02052) and the World Premier International Research Center Initiative (WPI), MEXT, Japan. J.W.A. and M.S.A. acknowledge support for CLAWS provided by the Office of the Under Secretary of Defense for Research and Engineering, Applied Research for the Advancement of S&T Priorities (ARAP) Program. F.J.G.A. acknowledges support from ERC (101141220-QUEFES) and the Spanish MICIU (PID2024-157421NB-I00 and CEX2024-001490-S).

**Author contributions:** D.N.B., L.N, S.Z. and Y.X. conceived the study. S.Z. conducted the nano-IR measurements. S.Z. built the nano-IR instruments used in this work. Y.X., supervised by L.N. and A.J.M., fabricated the devices used in the manuscript and investigated the transport behavior. D.S., supervised by C.D., fabricated some devices to investigate the tunneling. T.T. and K.W. grew the hBN crystals. J.Z., supervised by S.Z., provided theoretical modeling of tunneling. S.Z., D.N.B., L.N. and F.J.G.A. wrote the manuscript with input from all co-authors.

**Competing interests:** The authors declare that they have no competing interests.

**Data and materials availability:** All data are available in the main text or the supplementary materials.

**Supplementary Materials**

See the PDF file.

# Electrically driven plasmon-polaritonic bistability in Dirac electron tunneling transistors

Shuai Zhang *et al.*

Corresponding author emails: szhangphysics@gmail.com, lukas.novotny@ethz.ch, db3056@columbia.edu.

**This PDF file includes:**

    Materials and Methods
    Supplementary Text
    Figs. S1 to S13
    References

## Materials and Methods

### Device fabrication

Monolayer graphene flakes (natural graphite) and hBN flakes were exfoliated from bulk crystals onto SiO$_2$ (100 nm)/Si substrate at 65 °C and then identified from the corresponding optical contrast. The monolayer graphene flake was then cut into two halves using a probe tip (2.4 μm in diameter). The device was initially assembled using the standard dry transfer method following the sequence: top hBN, one half of graphene, hBN barrier, the other half of graphene, bottom hBN, with a polydimethylsiloxane (PDMS)/polycarbonate (PC) stamp. Here, two graphene flakes were picked up at 50°C and kept close to 0° twist angle, while hBN flakes were picked up at 80°C. The stack was then transferred onto a SiO$_2$ (285 nm)/Si substrate at 175°C. After removing the PC with chloroform, one-dimensional contacts (10 nm Cr/80 nm Au) were fabricated onto the stack using electron-beam lithography, reactive-ion etching, electron-beam evaporation, and metal lift-off. The surface of the top hBN in the device was then cleaned with atomic force microscope (AFM) contact mode.

### Nano-infrared scattering experiments

The nano-infrared scattering experiments were performed using a home-built cryogenic scattering-type scanning near-field optical microscope (s-SNOM) housed in an ultra-high vacuum chamber with a base pressure of ~$7 \times 10^{-11}$ torr. The s-SNOM is set up on a tapping-mode AFM. The tapping frequency and amplitude of the AFM are about 285 kHz and 70 nm, respectively.

The s-SNOM works by scattering tightly focused light from a sharp metallized AFM tip. The laser source is a wavelength-tunable quantum cascade laser from Daylight Solutions. The laser beam was focused onto the metallized AFM tip using a parabolic mirror with a 12-mm focal length. The back-scattered light was registered by a mercury cadmium telluride detector and demodulated following a pseudo-heterodyne scheme [60]. The signal was demodulated at the $n$th harmonic of the tapping frequency. To eliminate the far-field background, we chose $n$ = 3 and 4 in this work.

### Nano-photocurrent experiments

The nano-photocurrent measurements were performed simultaneously in the nano-IR imaging experiment. The laser power was set to be ~ 5 mW. The current was measured using a preamplifier with a gain setting of $10^5$ and a corresponding bandwidth larger than 1 MHz. In order to obtain the photocurrent generated by the near fields underneath the tip, the photocurrent was sent to a lock-in amplifier and demodulated at the second harmonic of the tip tapping frequency.

**Supplementary Note 1: Modeling of current-voltage curves**

**A. Electrostatic model**

The graphene/hBN/graphene heterostructure can be treated as a capacitor with both geometric and quantum capacitance. The applied electrostatic potential energy is distributed between the Fermi level shift $\Delta E_F$ and the Dirac point offset $\Phi$ (i.e., $eV_B = 2\Delta E_F + \Phi$), as shown in Fig. S1a. The Fermi level shift $\Delta E_F$ is defined as the difference between the Fermi level and the Dirac point, while the Dirac point offset $\Phi$ is defined as the energy difference between the Dirac points of the top and bottom graphene layers, that is, $\Delta E_F = |E_{F,T} - E_{D,T}| = |E_{D,B} - E_{F,B}|$ and $\Phi = |E_{D,T} - E_{D,B}|$. Figure S1(a) shows the band alignment for a negative bias voltage applied to the top electrode.

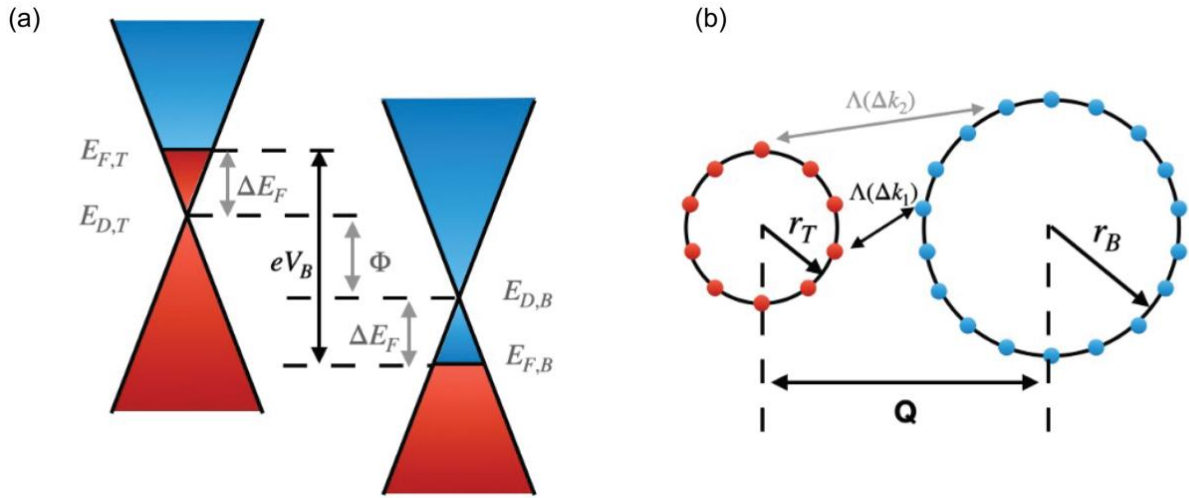

**Fig. S1. Energy band diagram of the resonant tunneling device.** (a) Schematic for band alignment in a graphene/hBN/graphene heterostructure where a negative bias voltage $V_B$ is applied to the top graphene electrode and the bottom graphene electrode is grounded. (b) Schematic of discretized states in the graphene electrodes at a specific energy level. Red and blue denote occupied and empty states, respectively.

The Dirac point offset $\Phi$ can be expressed in terms of the geometric capacitance of the heterostructure $C_{Geo}$ as

$$\Phi = e\left(\frac{en_{Gr}}{C_{Geo}}\right) = \frac{e^2 n_{Gr} d_{BN}}{\varepsilon_0 \varepsilon_{BN}} \quad (1)$$

where $n_{Gr}$ is the carrier density, $d_{BN}$ is the hBN thickness, and $\varepsilon_{BN} = 3.56$ is the electrical permittivity of hBN. To calculate the carrier density $n_{Gr}$, we start with the graphene density of state (DOS) $\rho_{gr}$,

$$\rho_{gr}(E) = \frac{g_s g_v |E|}{2\pi \hbar^2 v_F^2} \quad (2)$$

where the energy level $E$ is defined relative to the Dirac point, $g_s = 2$ and $g_v = 2$ are the spin and valley degeneracies, respectively, and $v_F \approx 1.15 \times 10^6$ m/s is the Fermi velocity in graphene. We can then obtain $n_{Gr}$ by integrating $\rho_{gr}$,

$$n_{Gr}(\Delta E_F) = \int_0^{\Delta E_F} \rho_{gr}(E) dE = \frac{\Delta E_F^2}{\pi \hbar^2 v_F^2} \quad (3)$$

and express the applied electrostatic potential energy as a function of $\Delta E_F$:

$$eV_B = \frac{e^2}{\pi \hbar^2 v_F^2} \frac{d_{BN}}{\varepsilon_0 \varepsilon_{BN}} \Delta E_F^2 + 2\Delta E_F \quad (4)$$

**B. Current-tunneling formalism**

The tunneling current is computed using the Bardeen transfer Hamiltonian approach [61],

$$I = g_v \frac{4\pi e}{\hbar} \sum_{\alpha,\beta} |M_{\alpha\beta}|^2 [f_T(E_\alpha) - f_B(E_\beta)] \delta(E_\alpha - E_\beta) \quad (5)$$

where $g_v$ is the valley degeneracy, and states in the top (T) and bottom (B) electrodes with energies $E_\alpha$ and $E_\beta$ are labeled by $\alpha$ and $\beta$, respectively. The $\delta$-function sets the condition for elastic tunneling. The tunneling matrix element

$$M_{\alpha\beta} = \frac{\hbar^2}{2m} \int dS \left( \Psi_\alpha^* \frac{d\Psi_\beta}{dz} - \Psi_\beta \frac{d\Psi_\alpha^*}{dz} \right) \quad (6)$$

dictates the transition probability. In this expression, $m$ is the free electron mass, while $\Psi_\alpha(r,z)$ and $\Psi_\beta(r,z)$ denote the wave functions on the top and bottom electrodes, respectively. The surface integral is evaluated at the middle point between the electrodes.

Applying suitable approximations [62] and labeling the states in the electrodes as $k_T$ and $k_B$, we evaluate the tunneling current as

$$I = \frac{8\pi e}{\hbar}\left(\frac{\hbar^2 \kappa e^{-\kappa d}}{2mD}\right)^2 \sum_{k_T, k_B}[f_T(E_{k_T}) - f_B(E_{k_B})]\delta(E_{k_T} - E_{k_B})\Lambda(\Delta k) \quad (7)$$

where $\kappa$ is the decay constant of the wave function in the barrier, $d$ is the separation between the electrodes, $D$ is a normalization constant for the $z$ part of the wave functions in graphene, and $\Delta k = k_T - k_B$ is the vector separating the top and bottom graphene electrode states in momentum space.

If the graphene layer has a finite area defined by $-\frac{L}{2} < x, y < \frac{L}{2}$, we can write

$$\Lambda(\Delta k) = \left|\frac{1}{A}\int_{-L/2}^{L/2} dx \int_{-L/2}^{L/2} dy\, e^{i\Delta k \cdot r}\right|^2 = \left|\text{sinc}\left(\frac{L\Delta k_x}{2}\right)\text{sinc}\left(\frac{L\Delta k_y}{2}\right)\right|^2 \quad (8)$$

where $\text{sinc}(x) = \sin(x)/x$, $A = L^2$ is the area of a perfectly crystalline graphene sheet, and $L = \sqrt{A}$ is the structural coherent length [62-64].

## C. Numerical simulations

To evaluate the elastic tunneling current at a given bias voltage, we integrate over the energy range that spans the Fermi levels of the two electrodes (in Fig. S1, from $E_{F,B}$ to $E_{F,T}$). At a given energy level, we exhaust all combinations of initial and final states and sum up their contributions. The schematic for discretizing states is shown in Fig. S1b. The Dirac cones are displaced by a misalignment vector $Q$ in momentum space due to the relative twist angle of the graphene sheets in real space. Because of the linear density of states near the Dirac point of graphene, and the fact that the energy level is farther from the Dirac point of the bottom electrode than from that of the top, more states are present on the bottom electrode. The sinc function decays rapidly away from the origin, and thus, states separated by $\Delta k_1$ make a more significant contribution to the tunneling current than those separated by $\Delta k_2$ in the general case. We simulate the tunneling current of a graphene/hBN/graphene heterostructure with 4 atomic layers of hBN for various misalignment angles. The results are shown in Fig. S2.

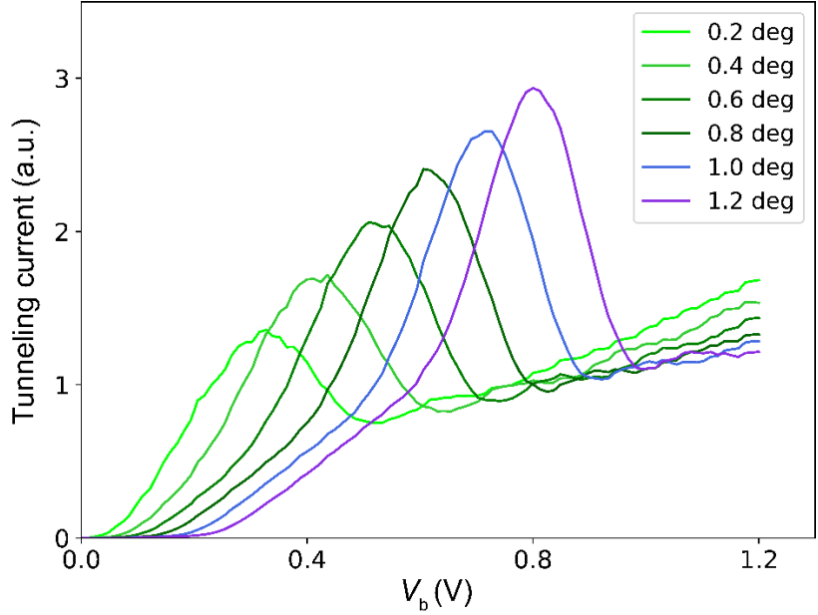

**Fig. S2. Simulated tunneling current as a function of applied bias voltage $V_b$ at twist angles of 0.2/0.4/0.6/0.8/1.0/1.2 degrees for a graphene/hBN/graphene heterostructure with 4 atomic layers of hBN**. Simulation performed assuming a temperature $T = 300K$ and a structural coherent length $L = 15$ nm.

We also simulated the I-V characteristics for a graphene/hBN/graphene heterostructure with 4 atomic layers of hBN, as was used in device 1 in our experiments. We obtained an excellent fit with a Pearson product-moment correlation of 0.98, as shown in Fig. S3.

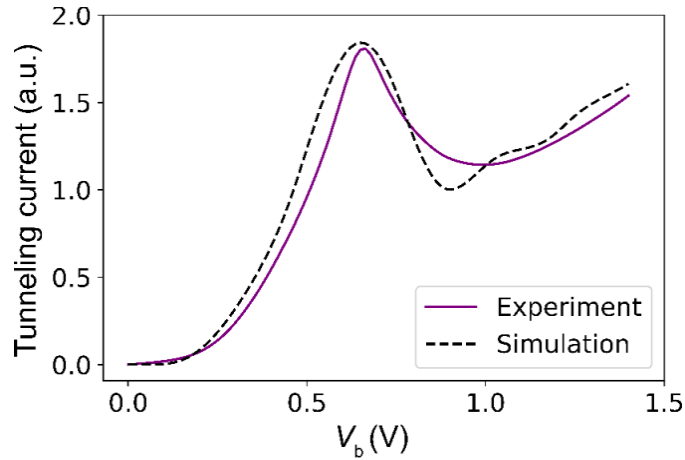

**Fig. S3. Comparison of simulated and measured I-V curves**. Measured I-V characteristics of a misaligned device with a 4-atomic-layer hBN barrier (purple) and corresponding fit with twist angle $\theta = 0.85°$ and structural coherent length $L = 11$ nm (black).

**Supplementary Note 2: Calculation of resonance peak position as a function of twist angle**

A negative differential conductance (NDC) peak appears in the I-V characteristics when the two Dirac cones intersect in a line. With the linear dispersion of graphene near the Dirac point, we can calculate the required bias voltage that would compensate for the misalignment momentum and, thus, give rise to the NDC resonance peak.

In the Brillouin zone, a Dirac cone located at the $\boldsymbol{K}$ point is at a distance of $\frac{4\pi}{3\sqrt{3}a}$ from the $\boldsymbol{\Gamma}$ point, where $a = 1.422$ Å is the carbon-carbon distance in graphene. For a twist angle $\theta$, the misalignment vector $\boldsymbol{Q}$ has a length $\frac{4\pi}{3\sqrt{3}a}\sqrt{2}\sqrt{1-\cos\theta}$. Resonance occurs for the Dirac point offset $\Phi_{I-peak} = \hbar v_F \|\boldsymbol{Q}\|$, such that the momentum misalignment is compensated for [9,11]. With the previously calculated relation between the Dirac point offset $\Phi$ and the Fermi level shift $\Delta E_F$, we can express the resonance peak position as a function of the twist angle as

$$eV_{I-peak} = \Phi_{I-peak} + 2\sqrt{\frac{\pi\hbar^2 v_F^2}{e^2}\frac{\varepsilon_0\varepsilon_{BN}}{d_{BN}}\Phi_{I-peak}} \quad (9)$$

We calculate the resonance peak position for various thicknesses of hBN in Fig. S4.

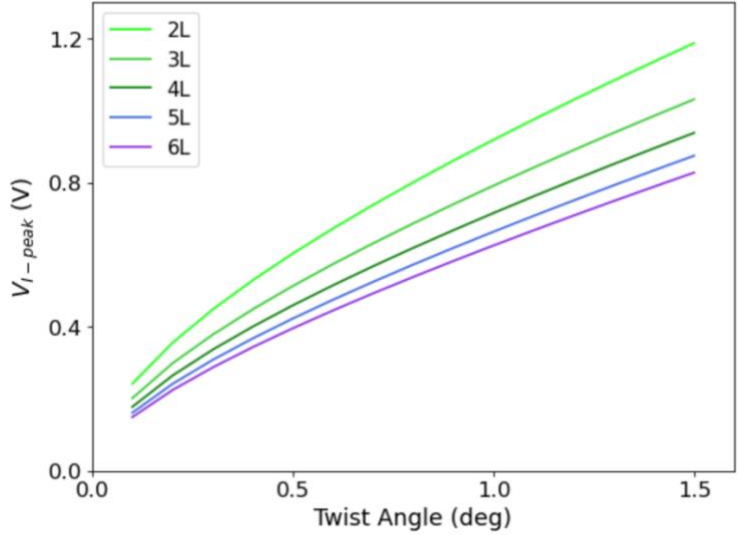

**Fig. S4. Resonance peak position in the *I-V* curves as a function of twist angle for a graphene/hBN/graphene heterostructure with 2/3/4/5/6 atomic layers of hBN.**

## Supplementary Note 3: Tunneling current bistability

The bistable I-V characteristics of the device are governed by the nonlinear equation $I_T(V_T) = I_L(V_b - V_T)$, where $I_T$, $I_L$, $V_T$, and $V_b$ are the tunneling current, the current across the load resistor, the voltage between the two graphene tunneling electrodes, and the voltage applied to the device, respectively. To simulate the bistable behavior of the circuit, we numerically calculate this nonlinear equation; the results are shown in Fig. S5. For bias voltages $V_b$ that yield three states, we ignore the state that is not stable and thus obtain the bistable I-V characteristics. By fitting the experimental data, we can also extract the contact resistance $R_c$. We find the contact resistance by maximizing the goodness of fit (GOF) of the simulation for multiple variable resistances $R_{\text{var}}$.

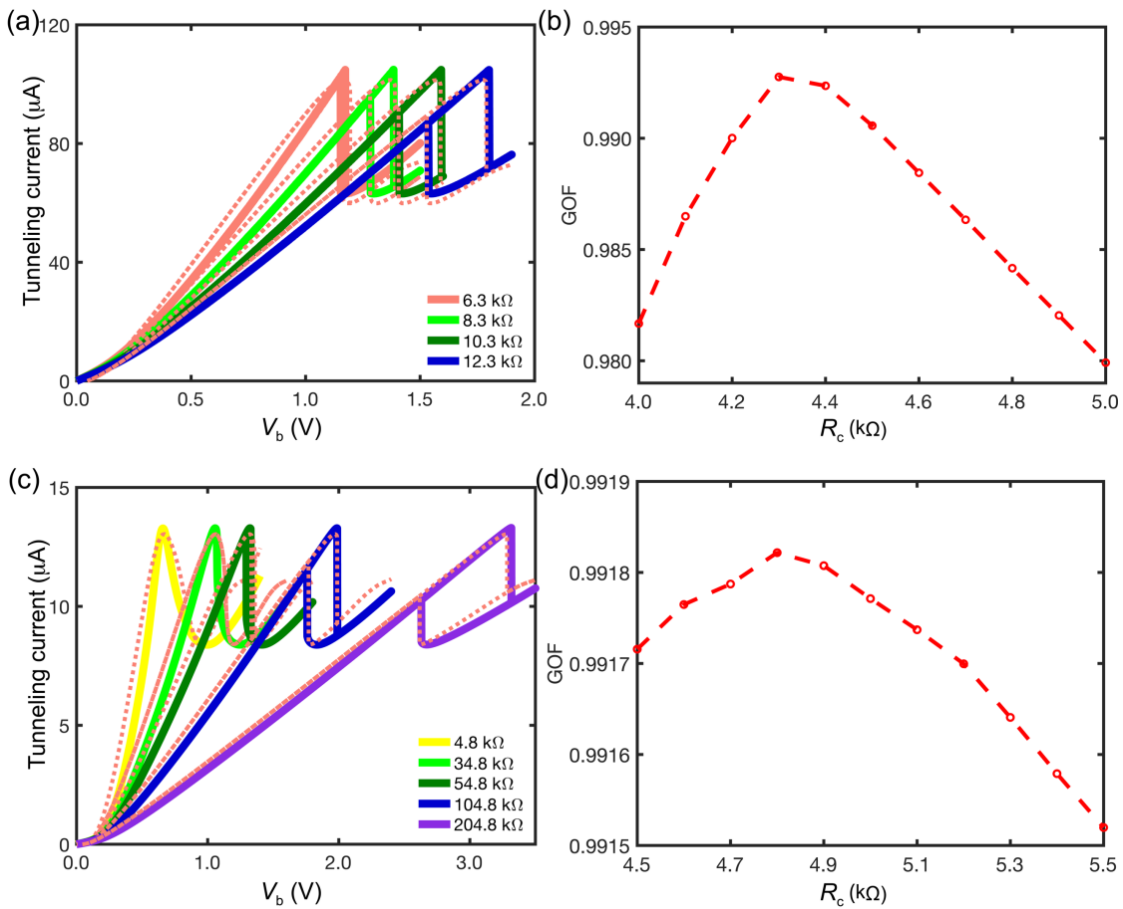

**Fig. S5. Tunneling current bistability simulations.** (a) Simulated (red dotted lines) and measured (solid lines) I-V characteristics for device 2 (D2) with various load resistance $R_L$. Added variable resistors $R_{\text{var}}$ have values 2 kΩ, 4 kΩ, 6 kΩ, and 8 kΩ. The contact resistance is extracted to be 4.3 kΩ. The I-V characteristic when no load resistance is placed in series is shown in Fig. 2(a). Simulation is done for a resonance tunneling junction with a 3-layer hBN barrier, twist angle $\theta =$

0.4 deg, and structural coherent length $L = 9.2$ nm. (b) Goodness of fit (GOF) of simulation as a function of contact resistance $R_c$. A maximum correlation of 0.993 is obtained for $R_c = 4.3$ kΩ. (c) Simulated (red dotted lines) and measured (solid lines) I-V characteristics for device 1 (D1) with various load resistances $R_L$. Added variable resistors $R_{var}$ have values 0, 30 kΩ, 50 kΩ, 100 kΩ, and 200 kΩ. The contact resistance is extracted to be 4.8 kΩ. Simulation is done for resonance tunneling junction with a 4-layer hBN barrier, twist angle $\theta = 0.7$ deg, and structural coherent length $L = 8.5$ nm. (d) Goodness of fit (GOF) of simulation as a function of contact resistance $R_c$. A maximum correlation of 0.992 is obtained for $R_c = 4.8$ kΩ.

**Supplementary Note 4: Plasmonic dispersion of the graphene resonant tunneling devices**

The double-layer graphene supports two plasmon modes: an optical mode and an acoustic mode. In the long-wavelength limit ($q \to 0$), the optical mode exhibits a dispersion $\omega_+(q \to 0) \propto \sqrt{q}$, and the acoustic mode has a dispersion $\omega_-(q \to 0) \propto q$ (ref. [21,65,66]). The calculated dispersions, represented by a 2D false color map of $\text{Im}(r_p)$, are shown in Fig S6. Figure S6(a and b) display the dispersions for devices with double-layer graphene and monolayer graphene, respectively. By comparison, an additional polariton branch appears in the dispersion of double-layer graphene. Figure. S6(c) illustrates the polariton momentum as a function of bias voltage across the tunnel junction. Within the experimental range of bias voltage, the momentum of the acoustic plasmon is significantly higher than that of nano-light imparted by the tip (denoted by a horizontal red dashed line) [67]. Consequently, only the optical mode can be effectively excited in our experiment. Figure S7 shows the nano-IR images acquired at varying voltages and the dispersion extracted from them via Fourier transformation. The experimental dispersion matches well with the calculated optical plasmon mode.

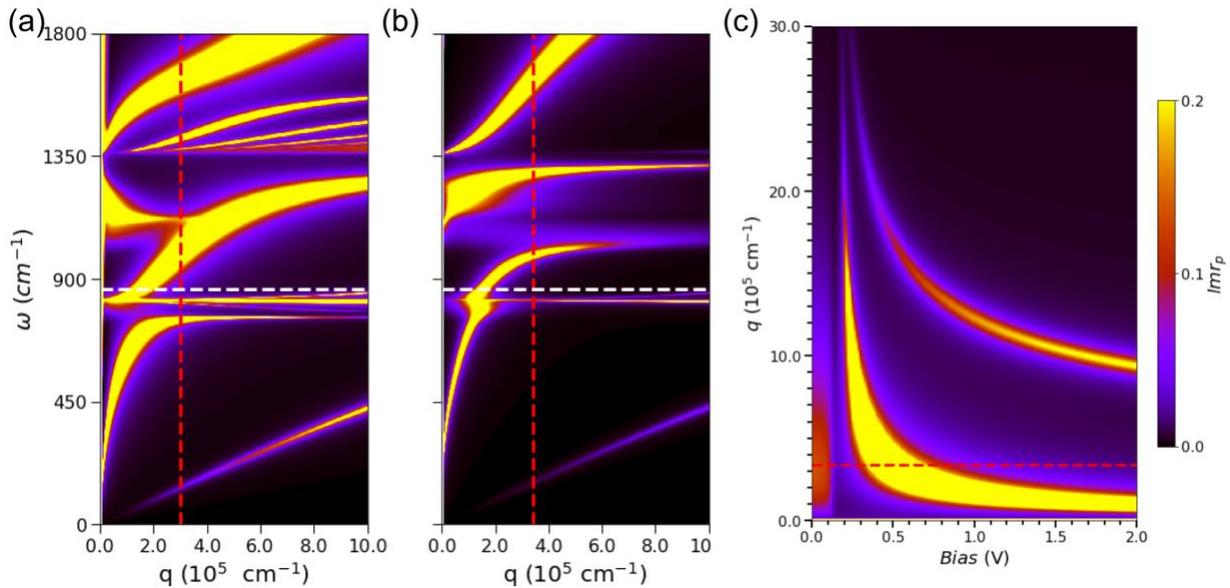

**Fig. S6. Calculated polaritonic dispersion**. (a) Calculated dispersion of polaritons in D1. The architecture of D1 is 2.1-nm hBN/graphene/1.3-nm hBN/graphene/35-nm hBN/285-nm SiO$_2$. The dispersion is visualized using the false color map of the so-called loss function (i.e., the imaginary part of the Fresnel reflection coefficient $\text{Im}(r_p)$). The vertical red dashed line indicates the momentum at which the tip-sample coupling reaches its maximum. The horizontal white dashed line denotes the energy at which the experiment is performed. (b) Calculated dispersion of a device

with the same structure as D1 except that there is only one graphene layer. In (a) and (b), the Fermi energy of graphene is set to 300 meV. (c) Plasmon-polariton momentum of D1 as a function of voltage between two graphene layers. There are two plasmon modes of acoustic and optical nature. The insets illustrate their corresponding electric field intensity profiles.

The selection of laser excitation frequency was guided by several factors, including polariton dispersion and momentum-matching conditions. Efficient excitation or probing of polaritons necessitates matching the momentum of the excitation structure to that of the polariton mode. In our case, we use the near field at the apex of an atomic force microscope tip to excite the polariton. The momentum imparted by the tip has a broad range and peaks at the momentum marked by a red vertical dashed line in Fig. S6. When the polariton momentum matches the tip momentum, the polariton can be efficiently excited. We used a laser with a frequency of 862 cm$^{-1}$, but it is not necessary to limit the excitation to 862 cm$^{-1}$ or a nearby frequency.

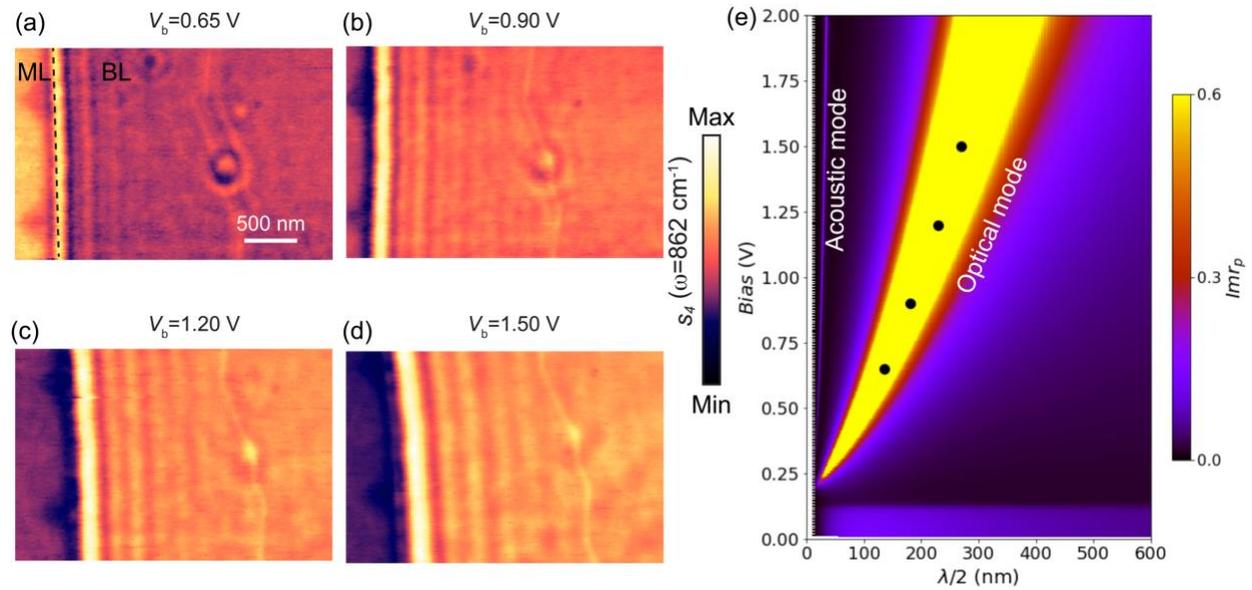

**Fig. S7. Bias-tuned plasmon polariton propagation.** (a-d) Nano-IR images obtained at varying bias voltages. All the data were taken on D1. The black dashed line in (a) denotes the boundary of the tunnel junction. (e) Measured plasmon-polariton dispersion derived from nano-IR images (dots), superimposed on the false color mapping of the calculated Im($r_p$). $\lambda/2$ denotes the half wavelength of the plasmon polaritons, and bias refers to the voltage across the tunnel junction.

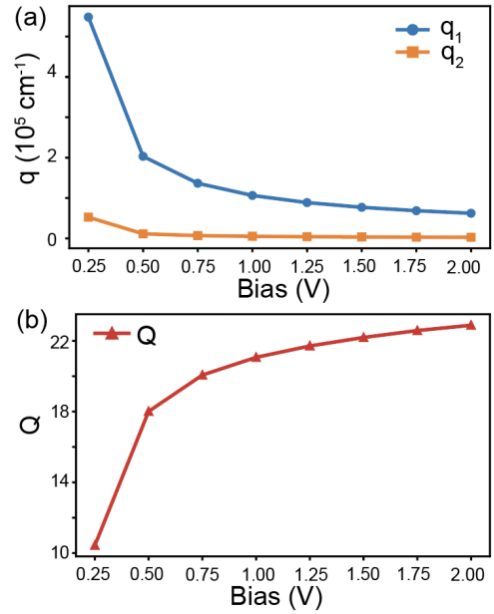

**Fig. S8. Plasmon-polaritonic bistability**. (a) The real and imaginary momentum of the optical mode plasmon polaritons as a function of bias voltage across the tunneling junction. The device architecture consists of 2.1-nm hBN/graphene/1.3-nm hBN/graphene/35-nm hBN/285-nm $SiO_2$, $\omega$=862 cm$^{-1}$. (b) Quality factor of the plasmon polaritons as a function of bias voltage.

To clarify the issue regarding the quality factor, we calculated the plasmon-polariton dispersion and quality factor as a function of bias voltage across the tunneling junction, as shown in Fig. S8. At low bias voltage, the graphene Fermi energy is low and thus strong interband transitions are allowed, resulting in enhanced Landau damping and low-quality factor. At higher bias voltage, the quality factor only slightly improves with the increase in bias voltage. In our devices, the state switch occurs at a voltage on the order of 1 V, where the quality factor remains nearly unchanged across the bistable transition.

## Supplementary Note 5: Plasmon-polaritonic bistability

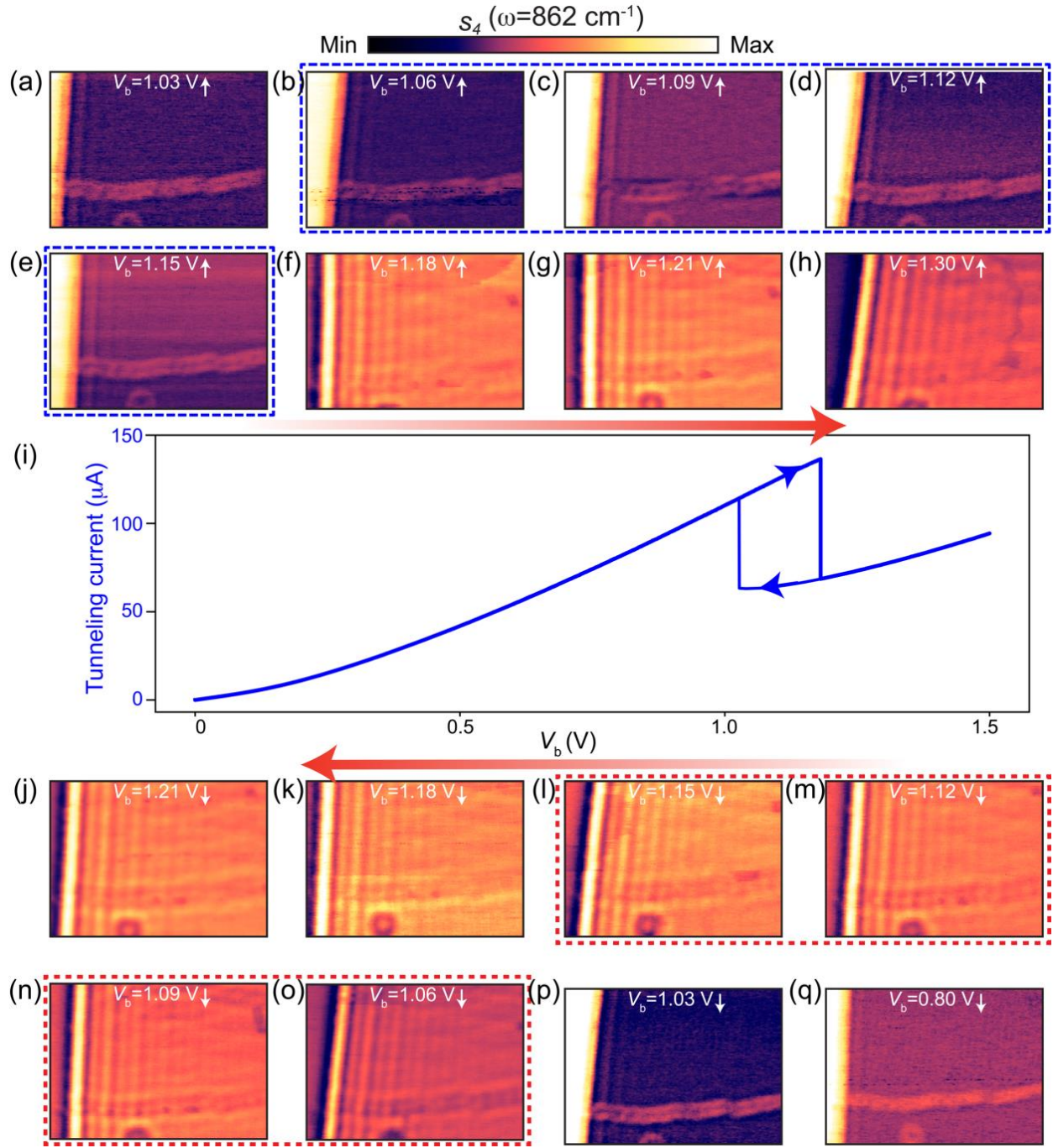

**Fig. S9. Plasmon-polaritonic bistability**. (a-h, j-q) Nano-IR images acquired when the bias voltage $V_b$ is swept upward (top panels) and downward (bottom panels). The images enclosed by red and blue lines show nano-IR images acquired at the same bias voltages, indicating that the plasmon depends on the bias-sweeping direction. (i) Bistable tunneling behavior revealed by the I-V curves obtained when sweeping $V_b$ in the direction of the arrows. All the data were acquired on device 2 (D2).

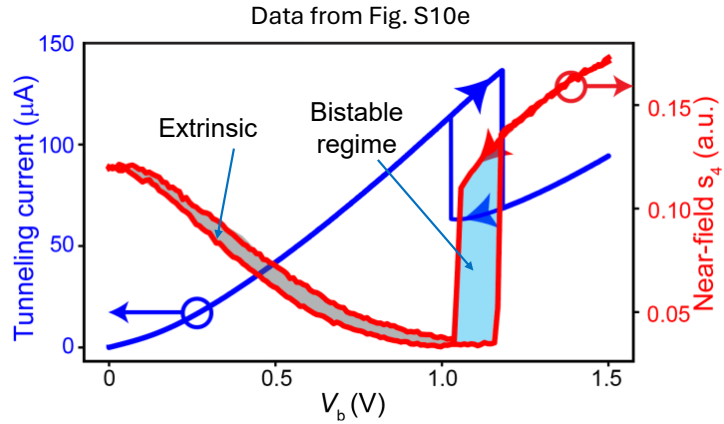

**Fig. S10. Correlation between bistable tunneling behavior (blue) and bistable plasmon-polaritonic response (red), both of which exhibit hysteretic behaviors.** The hysteresis shaded in cyan represents the plasmonic bistable regime, whereas the hysteresis shaded in gray arises from extrinsic effects, as described in the text.

We first differentiate between the two types of hysteresis shown in Fig. S10. The hysteresis arising from intrinsic bistability in this study is pronounced and always appears concomitantly with the sharp state-switch. In contrast, extrinsic hysteresis is minor and, in some cases, barely discernible.

The hysteresis curves are extracted from 2D nano-IR mapping, the two axes of which represent bias voltage and position. Acquiring a complete 2D mapping takes approximately two hours; any drift in the sample position or the optical alignment during this time will result in discrepancies in the IR signals between the forward and backward bias scans. As a result, extrinsic hysteresis with respect to bias voltage is formed.

The switching time of the bistable states is determined by the RC time constant, where R and C denote the resistance and the capacitance of the device, respectively. In our device, the capacitance originates from the tunneling junction, while the resistance includes contributions from junction resistance itself and external components, such as the graphene-metal contact resistance and the tunable series resistor. To speed up the device response, we should minimize both R and C.

The tunneling resistance exponentially decreases when the tunneling barrier thickness is thinned down, while the capacitance linearly increases. Hence, optimizing the switching speed requires balancing these two factors. Ideally, the tunneling barrier should approach the monolayer limit of van der Waals material. In principle, the device footprint does not influence the response time since R and C scale inversely. However, in practice, miniaturization increases the contact resistance because the contact area decreases.

To summarize, to extend the operating frequency and improve switching performance, the following optimization strategies can be pursued: (a) Reducing the tunneling barrier thickness. (b) Selecting tunneling barrier material with an appropriate bandgap and dielectric function, as these parameters determine the resistance and the capacitance. (c) Improving the contact fabrication to minimize the contact resistance. (d) Optimizing the thickness of graphene. Few-layer graphene reduces the tunneling resistance; however, it may compromise the resonant tunneling peak-to-valley ratio (PVR) and the tunability of plasmon-polariton.

Experimentally demonstrating the ultimate switching time faces several technical challenges. First, fabricating devices with a monolayer tunneling barrier is a formidable task because even a nanoscale crack created during stacking can short graphene layers and destroy the tunneling effect. Second, when the tunneling barrier reach the monolayer limit, the contact resistance becomes comparable to or even larger than the tunneling resistance, thereby dominating the total resistance and limiting the achievable speed. With the advancement of device fabrication, these challenges would be overcome, enabling a meaningful experimental demonstration of ultrafast switching time.

We envision that the device footprint can be reduced to $10 \times 10$ nm$^2$ (Ref. [46,68]). Based on the known device parameters, detailed below, the device can operate at the single-electron level. In other words, the plasmon-polaritonic state can be switched by the injection or ejection of a single electron, and only a few electrons are required to store the information or present the device status.

Details of the above estimation are shown here. The dielectric constant of BN is 3.5; the tunneling barrier is 5-layer BN; device footprint is $10 \times 10$ nm$^2$; the bias voltage between the two switchable states is around 0.1 V. Calculated capacitance of device is $1.9 \times 10^{-18}$ F. The number of charges used for state-switch is $N = \frac{CV}{e} = \frac{1.9 \times 10^{-18} \text{F} \times 0.1 \text{ V}}{1.6 \times 10^{-19} \text{C}} = 1.2$.

**Supplementary Note 6: Load-resistance-engineered plasmon-polaritonic bistability**

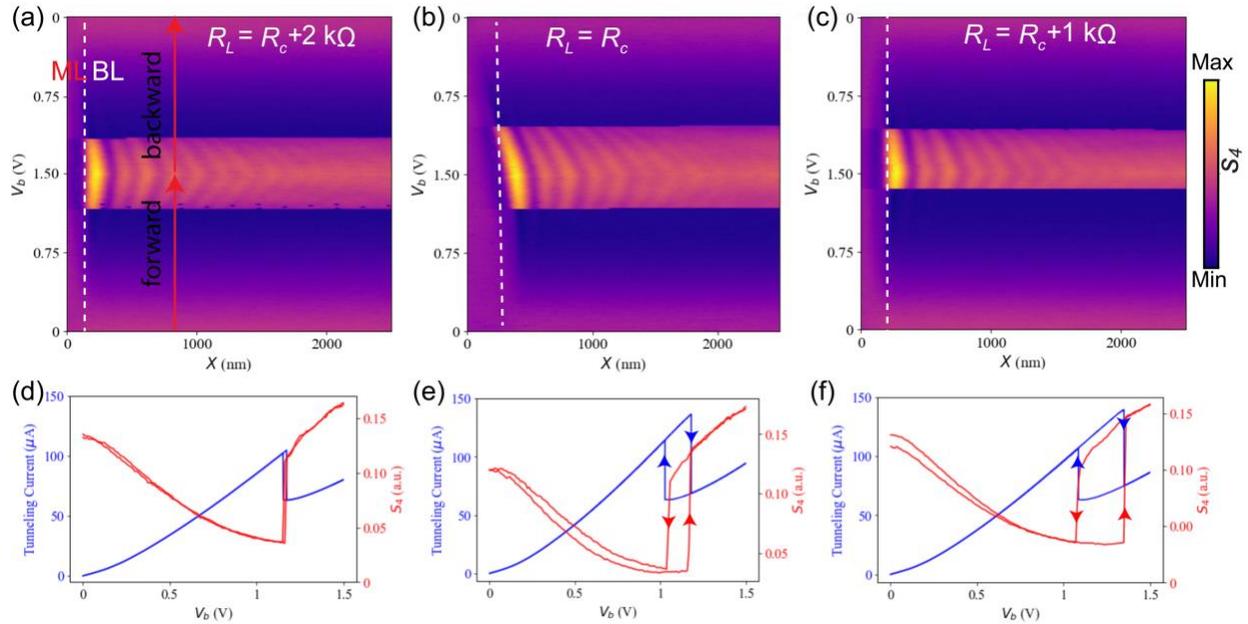

**Fig. S11. Plasmon-polaritonic bistability tuned via load resistance for device 2.** (a, b, and c) Nano-IR scattering ($s_4$) line scans along a line perpendicular to the boundary between the graphene monolayer and bilayer for various load resistances, as indicated in the upper labels. In the course of the measurement, the tip was repeatedly scanned along the same line while the bias voltage was swept from zero to maximum and then swept back to zero. The vertical white dashed line marks the boundary between the graphene monolayer (ML) and bilayer (BL). (d, e, and f) Line profiles of $s_4$ (red) extracted from the corresponding nano-IR scattering line scans. The scanning lines are indicated by two red lines in (a), which denote upward and downward bias voltage scans. The tunneling current traces (blue) are also plotted for comparison. These data demonstrate that the emergence of plasmon-polaritonic hysteresis and the width of the hysteresis loop can be tuned by load resistance.

**Supplementary Note 7: Plasmon-polaritonic bistability-engineered photocurrent**

The plasmonic bistability is expected to translate into analogous behavior of the photocurrent response. We performed nano-photocurrent measurements to investigate the plasmon-engineered photocurrent [45,69,70]. In graphene, the photocurrent is dominated by the photo-thermoelectric effect [71,72]. The electron temperature in graphene can be efficiently increased via the power absorbed from propagating plasmons. This increase in electron temperature can generate a photocurrent. Therefore, the plasmonic bistability and the photocurrent bistability occur concomitantly. Figure S11 shows the anticipated photocurrent bistability. The plasmonic fringes are observed in both nano-IR scattering and nano-photocurrent images.

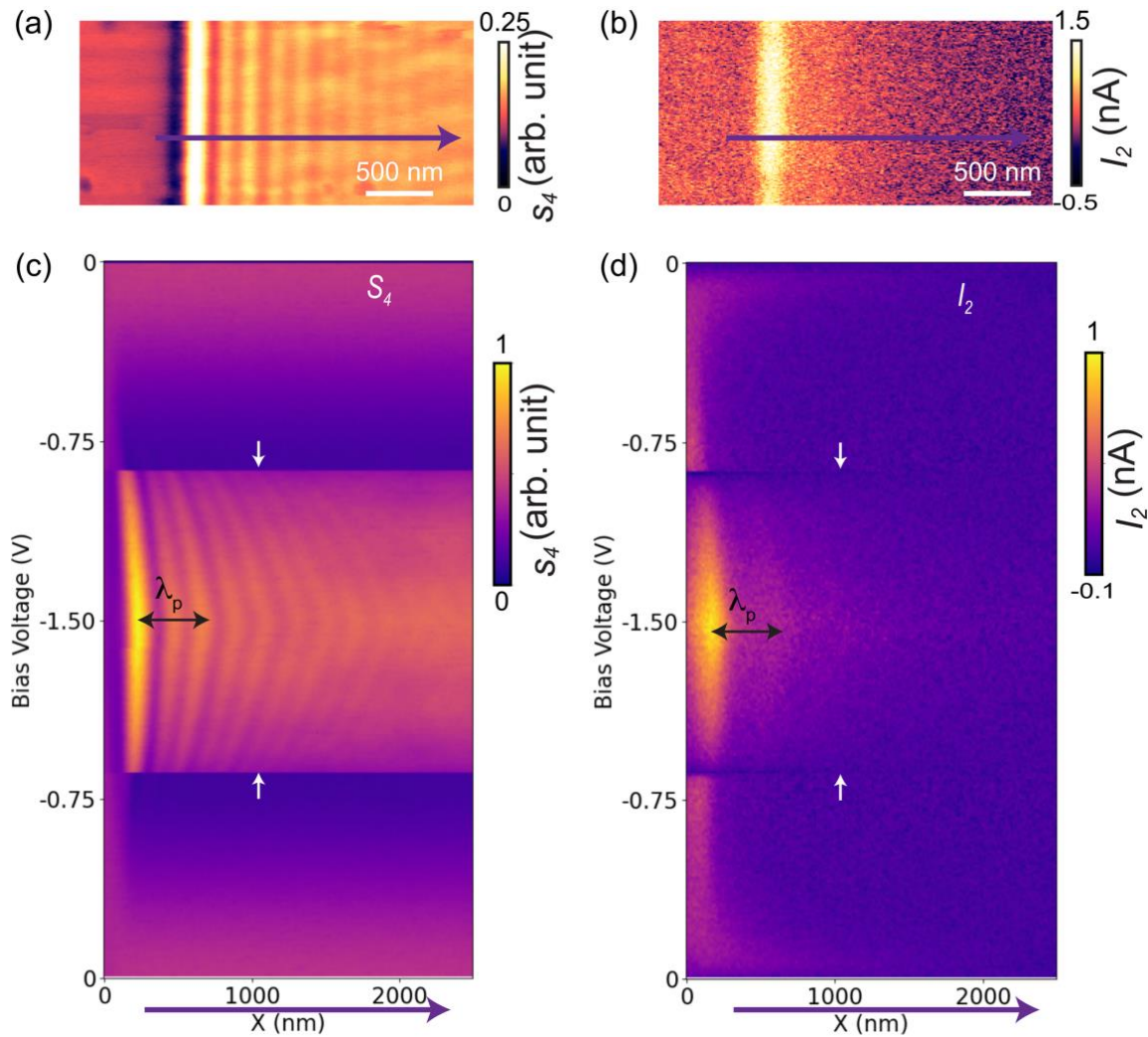

**Fig. S12. Bistable photothermoelectric current.** (a) Nano-IR image acquired at $V_b = -1.5$ V. (b) Nano-photocurrent taken simultaneously with (a). (c) Bistable plasmon-polaritonic response observed in the 2D nano-IR scattering data, $s_4$. The white arrows denote the abrupt switch between polaritonic states. In the course of the measurement, the tip was repeatedly scanned along the same line (purple lines in (a) and (b)) while the bias voltage was swept from zero to -1.5 V and then swept back to zero. (d) Bistable photocurrent arising from the plasmon-polaritonic response. Fringe patterns with plasmon polariton period $\lambda_p$, indicated by a black arrow, can be observed. The photocurrent response abruptly switches to another state at the voltage where the plasmon polariton switches, indicated by white arrows. All the data were acquired on D2.

**Supplementary Note 8: Optical bistability and the nature of plasmon-polaritonic bistability**

Optical bistability incorporates four ingredients, namely, optical input, optical output, nonlinearity, and the driving action/control. All kinds of optical bistability have optical input and output, but they are rooted in various mechanisms of nonlinearity and are correspondingly driven by different types of actions and controls. In the following, we elaborate on these points to illustrate the more generalized concept of optical bistability as it has been understood since its discovery.

1) "Conventional" optical bistability and "generalized" optical bistability

Optical bistability was first discovered in sodium vapor [73] and semiconductors [74]. These bistable states arise from optical nonlinearity, such as the nonlinear Kerr effect, and are driven by light illumination. All these scenarios are categorized as all-optical bistability or conventional optical bistability.

In the past decades, optical bistability has been studied in a much broader context. While the input and output are optical, the nonlinearity and driving are not constrained to be optical, as detailed below. Notably, the optical outputs have also been extended to broader concepts. The output is not always light reflection/transmission, and it can involve hybrid light and matter, such as exciton-polaritons [3,39] and plasmon polaritons [34].

2) The origin of the nonlinearity

   i. Optical nonlinearity

Optical nonlinearities, such as the Kerr effect, are pivotal in achieving optical bistability. The Kerr effect induces an intensity-dependent refractive index change, which can modify the resonance conditions in optical cavities, leading to bistable behavior [33].

   ii. Mechanical nonlinearity

In artificial nanostructures with coupling of optical and mechanical subsystems, driving the nonlinear mechanical response can result in optical bistability [36].

   iii. Phase transition

Phase-change materials have been widely used for rewriteable optical data storage, for example, compact discs (CD) and digital versatile discs (DVD) [37]. The bistable states depend on the distinct phases of the data storage materials. In addition, bistability based on phase-change materials is usually nonvolatile.

    iv.    Electrical nonlinearity

When light is strongly coupled with elementary excitations of matter, polaritons are formed. The half-light and half-matter nature of the polariton facilitates the electrical driving of the bistability. Electrical field-driven bistable exciton polaritons were demonstrated in a quantum well in a cavity[3,39]. One of the underlying trigger mechanisms of bistability is the nonlinear dependence of electron and hole tunneling on the electric field. The tunneling process tailors the exciton lifetime, thereby resulting in bistable exciton-polaritons [3].

3) The control of optical bistability

    i.    Light

The optical bistability arising from optical nonlinearity is driven by the light illumination. To reduce the required electric field, a cavity or resonator can be employed [75]. It is worth noting that optical frequency detuning can also be used to trigger optical bistability, which was demonstrated in a Rydberg vapor [76].

    ii.    Heat and optical heating

For the bistability relying on phase transition, the states are usually driven by thermal effect. The heat of absorbed light is commonly used to drive the phase transitions, thus providing the necessary optical properties for optical bistability [77].

By exploiting the coupling of optical and mechanical subsystem, the optical bistability can be driven by mechanical nonlinearity. The optical thermal effect has been used to generate mechanical bistability and, thus, optical bistability [36].

    iii.    Force (e.g., optical force)

The forces produced by acoustic excitations and optical pressure have been used to generate optical bistability in nano-machined resonators. The underlying mechanism is the deformation-induced electric field distribution [78] or the reconfigurable plasmon response via mechanical deformation [35].

    iv.    Electricity

Polaritonic bistability is usually driven by electric stimuli. An extensively investigated system is bistable exciton-polaritonic states in semiconductor quantum wells in a distributed Bragg reflector (DBR) cavity. The bistable state switching is driven by current [39] or electric field [3].

Plasmon polaritons are hybrid excitations of light and matter – specifically, mobile electrons. While established literature has indeed usually exploited optical nonlinearity to demonstrate bistability, we instead leverage the matter part (the electrons), that is, carrier density tuned by resonant tunneling. The plasmon polaritonic bistability, the first of its kind, is achieved by designing a specific double-layer system to take advantage of the strong nonlinearity of tunneling electrons and the superior plasmon in graphene. To avoid the ambiguity in terminology, we emphasize that our work has demonstrated electrically driven plasmon-polaritonic bistability.

**Supplementary Note 9: The role of plasmon polaritons in bistability**

Bistable plasmon-polaritonic response is in stark contrast to its "conventional" optical counterparts because plasmon polaritons are a hybridization of light with matter, which enable harnessing and manipulation of light at the nanometer scale. Therefore, plasmon polariton-based circuits can merge electronics and photonics at the nanoscale [49]. The enhanced field of plasmon polariton can be leveraged in the sensing and readout of plasmonic states. The small footprint enabled by the nanoscale plasmon also improves energy efficiency. Since polaritons possess appealing characteristics, as detailed below, we would conclude that from a device perspective, the appealing features reside in transport properties as well as optical performance.

The role of plasmon polaritons in this work and its appealing features are detailed in the following.

(1) High spatial confinement facilitates integration.

Whereas optical fields often suffer from diffraction limits and require complex nanophotonic structures for localization, the electric fields associated with plasmons can be confined to nanoscale regions [43]. Plasmonics based on plasmon polaritons demonstrated here compress the light to subwavelength regions, and its "killer application" is to be implemented as digital highways.

Electronic circuits, with a dimension of tens of nanometers, enable controlling the transport and storage of electrons. However, electrical circuits cannot satisfy the requirements of high-speed digital information communications. This constraint can be solved by optical connections based on optical fibers and photonic circuits. Unfortunately, photonic components are bulky (in the micrometer scale), which has limited the combination of electronics and photonic components within the same chip. However, plasmon polariton-based circuits can merge electronics and photonics at the nanoscale [49].

(2) An enhanced electric field promotes sensing and its electrical readout.

Along with the extreme spatial confinement, graphene plasmon polaritons strongly enhance the electric field, boosting the photocurrent response. In this study, we leverage this enhancement to achieve bistable photocurrent responses, where the plasmonic states modulate optoelectronic behaviors. Moreover, plasmon polaritons can control the optoelectronic response, while the photocurrent works as a readout/sensor out of the plasmonic states.

In addition to acting as a sensor of photons, plasmon polariton can be used for biosensing [79]. Furthermore, the plasmon polariton has a reduced mode volume that efficiently tailors the emission lifetime [80].

(3) Improved energy efficiency

Reading of the plasmonic states could be performed either electrically or optically/plasmonically. In this respect, a relevant question for future studies relates to the minimum device area required to sustain a robust plasmonic bistability behavior. We would expect the smallest footprint to be limited in lateral size to the in-plane plasmon wavelength of ~10 nm, which reduces the number of electrons per bit. As a result, the plasmon polaritons enable a substantial saving in power.

**Supplementary Note 10: Perspective on the application of plasmon-polaritonic bistability**

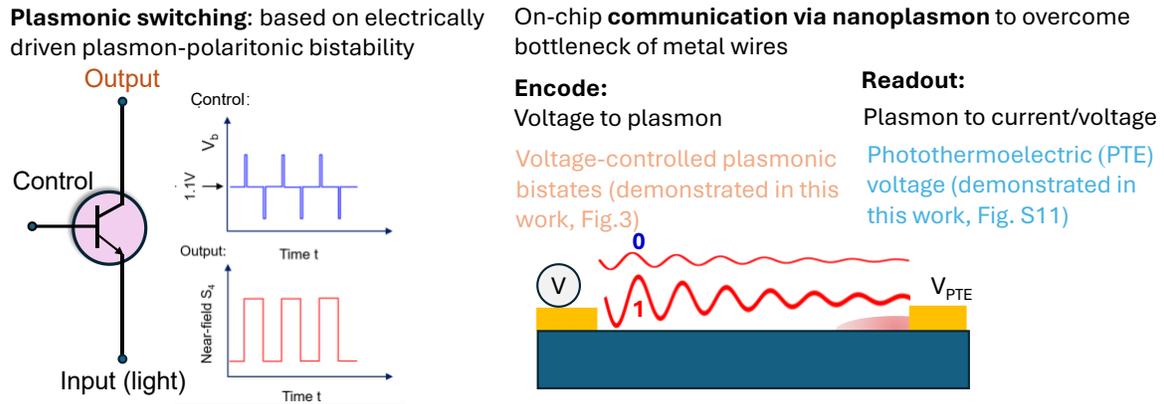

**Fig. S13. Applications of electrically driven plasmon polaritonic bistability.** Left: plasmonic switching/memory; Right: optical interconnections via voltage-controlled nano-plasmonic bistability.

We now provide examples of how a bistable plasmonic system could be utilized in optically read memory, nonlinear optical switching, and plasmonic interconnects.

(a)     Plasmonic memory/switches/modulators

Bistable plasmonic states enable the construction of an optoelectronic flip-flop memory unit, as proposed in Fig. S13. Based on the parameters of D2 in Fig.3 of the main text, when the junction is biased at 1.1 V, a positive voltage pulse triggers the output near-field signal to high intensity. The output stores "1" and holds it until a negative voltage pulse resets the status to "0". Such a flip-flop device stores the electrical information and converts the electrical signal to an optical signal as an interlink.

(b)     Optical interconnections via voltage-controlled nano-plasmonic bistability

In modern nanophotonic and optoelectronic circuits, achieving efficient, high-speed, and low-power optical interconnects is crucial. Nano-plasmonic bistability, in which the system exhibits two stable states under the same external conditions, offers a promising way to enable such interconnections at the nanoscale.

By leveraging voltage-controlled nano-plasmon polaritons (NPPs), signals can be dynamically switched, routed, and transferred between different functional blocks in an integrated system, as illustrated in the right panel of Fig. S13.

**Supplementary References**